\documentstyle[fullpage,preprint]{acmconf}
\input{psfig}
\newtheorem{example}{Example}[section]
\newtheorem{lemma}{Lemma}

\newtheorem{theorem}{Theorem}
\newtheorem{definition}{Definition}
\newtheorem{algorithm}{Algorithm}
\newcommand {\qed}{\vrule height4pt width4pt depth2pt}

\begin{document}
\bibliographystyle{alpha}

\title{\LARGE\bf Similarity-Based Queries for Time Series Data}
\author{{\bf Davood Rafiei}\\
   	{\small drafiei@db.toronto.edu}\\
   	{\small Department of Computer Science}\\
   	{\small University of Toronto} 
   \and {\bf Alberto Mendelzon}\\
   	{\small mendel@db.toronto.edu}\\
   	{\small Department of Computer Science}\\
   	{\small University of Toronto}}
\date{}

\setcounter{page}{13}
\toappear{\bf In Proceedings of the ACM SIGMOD International Conference
on Management of Data, May 1997, Tucson, Arizona}

\maketitle

\begin{abstract}
We study a set of linear transformations on the Fourier series
representation of a sequence that can be used as the basis
for similarity queries on time-series data.
We show that our set of transformations
is rich enough to formulate operations such as moving average
and time warping. We present a query processing algorithm
that uses the underlying R-tree index of a multidimensional data set
to answer similarity queries efficiently. Our experiments show that
the performance of this algorithm is competitive to that of processing
ordinary (exact match) queries using the index, and much faster than
sequential scanning. We relate our transformations to the general
framework for similarity queries of Jagadish et al.

\end{abstract}

\section{Introduction}
Time-series data are of growing importance in many new database
applications, such as data mining or warehousing.
A time series is a sequence of real numbers,
each number representing a value at a time point.
For example, the sequence could represent stock or commodity prices,
sales, exchange rates,
weather data,
bio\-me\-dic\-al
measurements, etc.
We are often interested in similarity queries on time-series data
\cite{shape95,Ag+95}.  For example, we may want to find stocks that behave in approximately the same way (or approximately the opposite way, for hedging); or stocks that increased linearly up to October 1987, and then crashed; or years when the temperature patterns in two regions of the world were similar.
In this type of queries, approximate rather than exact matching is required.

A naive approach is to compute the Euclidean distance (or any other distance,
such as the city-block distance) between two objects (in general) or two
time sequences (in particular), and call two sequences 
similar if their distance is less than a user-defined threshold. 
Time sequences are usually long, so the distance computation can
be time consuming. A solution is to map time sequences into
the frequency domain using the Fourier transform, and use the first few 
coefficients to filter out non-similar data. This has the advantage
that
spatial indexing techniques can be used to index
time sequences by viewing them as tuples of Fourier
series coefficients,
that is, points in a low-dimensional space \cite{AgFaSw93,FRM94}.

A problem with this approach is that the user has no control 
over the meaning of similarity other than providing a threshold.
There are many similarity queries that such a fixed predefined notion of similarity fails to capture; for example, one may consider two stocks similar if they have almost the same price fluctuations, even though one stock might sell twice as much as the other. Consider the following
motivating examples:

\begin{figure}[htbp]
\psfig{figure=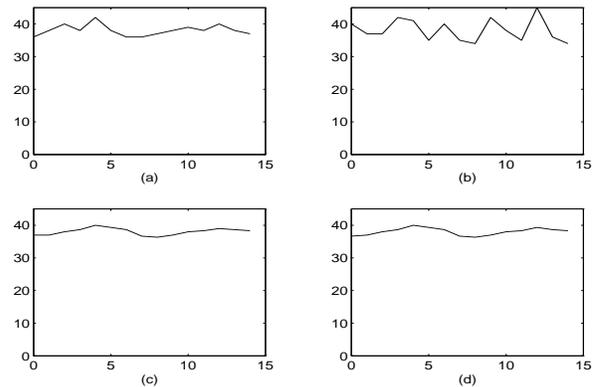,height=2in,width=3.05in}
\caption{(a) Time sequence 
$\vec{s_1}=$ (36,38,40,38,42,38,36, 36,37,38,39,38,40,38,37),
(b) time sequence $\vec{s_2}=$ (40,37,37,42,41,35,40,35,34,42,38,35,45,36,34),
(c) the 3-day moving average of $\vec{s_1}$, and
(d) the 3-day moving average of $\vec{s_2}$}
\label{fig:mavg}
\end{figure}

\begin{example}{\rm 
\label{example:1}
Suppose $\vec{s_1}=$ (36,38,40,38,42,38,36,36, 37,38,39,38,40,38,37) and
$\vec{s_2}=$ (40,37,37,42,41,35,40,35, 34,42,38,35,45,36,34) are two time
sequences that correspond to the closing prices of two stocks.
Looking at Figure~\ref{fig:mavg}(a),(b), the sequences do not
appear very similar.
This is justified by the high Euclidean distance
$D(\vec{s_1},\vec{s_2})=11.92$  between them. However, 
if we look at the
three-day moving averages of the two sequences (Figure~\ref{fig:mavg} (c),(d)),
they do look quite similar. The Euclidean
distance between the three-day moving averages of two sequences is
$0.47$.

Moving averages are widely used in stock data analysis 
(for example, see \cite{Stockanal}).
Their primary use is to
smooth out short term fluctuations and depict the underlying
trend of a stock. The computation is simple; the $l$-day moving average
of a sequence $\vec{s}=(v_1,\ldots,v_n)$ is computed as follows: 
the mean is computed for an $l$-day-wide window placed over the 
end of the sequence; this will give the moving average
for day $n - \lfloor l/2 \rfloor$; the subsequent values are obtained 
by stepping the window 
through the beginning of the sequence, one day at a time. 
This will produce a moving average of length $n-l+1$. 
We use a slightly different version of moving average which is 
easier to compute in our framework. We circulate the window to the
end of the sequence when it reaches the beginning. 
This gives us a moving average of length $n$.
It turns out when the length of the window is small enough compared to
the length of the sequence, which is usually the case in practice, both 
averages are almost the same.
}\end{example}

\begin{figure}[htbp]
\psfig{figure=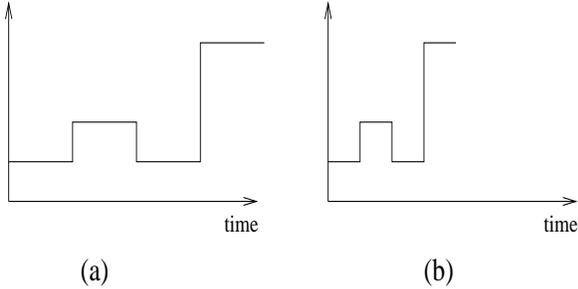,height=1.5in,width=3in}
\caption{(a)Time sequence $\vec{s}=(20,20,21,21,20,20,23,23)$
         (b)time sequence $\vec{p}=(20,21,20,23)$}
\label{fig:expan}
\end{figure}

\begin{example} 
\label{example:2}{\rm
Consider two time sequences
$\vec{s}=$ (20,21 ,21,21,20,21,23,23) and
$\vec{p}=(20,21,20,23)$ that are sampled with different frequencies 
(Figure~\ref{fig:expan}).
For example, $\vec{s}$ could be the closing price of a stock taken every day,
and $\vec{p}$ could be the closing price of another stock taken 
every other day. 
A typical query is ``is $\vec{p}$ similar to $\vec{s}$ ?''.
The sequence $\vec{s}$ is twice as long as $\vec{p}$, 
so they cannot be compared directly.
The Euclidean distance between $\vec{p}$ and any subsequence of length
four of $\vec{s}$ is more than $1.41$. 
If the time dimension of $\vec{p}$ is scaled by 2, i.e., every value
``$v_i$'' is replaced by ``$v_i,v_i$'',
the resulting sequence will be identical to $\vec{s}$. 
This operation is usually called {\it time warping} 
(for example, see \cite{SanKru83}).
}\end{example}

We propose a class of transformations that can be used in
a query language to express similarity in a fairly general way,
handling cases like the two examples above.
Given an R-tree index \cite{Gutt84} constructed 
on a data set,
we describe a fast query processing algorithm 
that uses the index to filter out unrelated data from the answer set of
a similarity query. For example, we demonstrate that an index
structure for moving average can be constructed on the fly
from the existing index, and it can be used to speed up
the query processing. 
We show that this approach not only is faster than sequential
scanning, but also introduces no extra disk overhead. 
To the best of our knowledge, this is the first indexing method
that can handle moving average and time warping in the context of
similarity queries.

The organization of the rest of the paper is as follows:
The rest of the current section provides some basic material 
on the discrete Fourier transform and a survey of related work. 
Section~\ref{sec:examples} motivates the work by discussing
possible applications
to stock data analysis.
Our definition of similarity queries is discussed in Section~\ref{sec:simq}. 
In section~\ref{sec:index} we develop an indexing method for these
similarity 
queries. 
Section~\ref{sec:exper} presents experimental performance results.
We conclude in Section~\ref{sec:conclude}.

\subsection{The Discrete Fourier Transform}
\label{sec:DFT}
In this section, we briefly review the Discrete Fourier Transform (DFT) and its
properties. Let a time sequence be a finite duration
signal $\vec{x}=[x_t]$ for $t=0,1, \cdots ,n-1$. The DFT
of $\vec{x}$, denoted by $\vec{X}$, is given by 
\begin{equation}
X_f = \frac{1}{\sqrt{n}}\sum_{t=0}^{n-1} x_t e^{\frac{-j2\pi tf}{n}} \ \ \
f=0,1, \cdots, n-1
\label{eq:fft}
\end{equation}
where $j=\sqrt{-1}$ is the imaginary unit.
Throughout this paper, unless it is stated otherwise, we use small letters 
for sequences in the
time domain and capital letters for sequences in the frequency domain.
The inverse Fourier transform of $\vec{X}$ gives the original signal, i.e.,
\begin{equation}
x_t = \frac{1}{\sqrt{n}}\sum_{f=0}^{n-1} X_f e^{\frac{j2\pi tf}{n}} \ \ \
t=0,1, \cdots, n-1
\label{eq:ifft}
\end{equation}
Following the convention of \cite{AgFaSw93,FRM94}, we have $1/\sqrt{n}$ 
in the front of both Equations~\ref{eq:fft} and \ref{eq:ifft}. 
The energy of signal $\vec{x}$ is given by
the expression
\begin{equation}
\label{eq:energy}
E(\vec{x}) = \sum_{t=0}^{n-1} |x_t|^2  .
\end{equation}
The convolution of two signals $\vec{x}$ and $\vec{y}$ is given by
\begin{equation}
\label{eq:conv}
Conv(\vec{x},\vec{y})_i = \sum_{k=0}^{n-1} x_k y_{i-k} \ \ 
i=0,1, \cdots,n-1
\end{equation}
where $i-k$ is computed modulo $n$.
This convolution is usually called {\it circular convolution}.
Equations~\ref{eq:energy} and \ref{eq:conv} are unchanged in
the frequency domain.

The following properties of DFT can be found
in any signal processing textbook (for example, see \cite{Oppenheim75}). 
The symbol $\Leftrightarrow$ denotes a DFT pair.
\begin{description}
\item[Linearity] 
 \begin{equation}
 a\vec{x} + b\vec{y} \ \Leftrightarrow a\vec{X} + b\vec{Y}
 \label{eq:linear}
 \end{equation}
 for arbitrary constants $a$ and $b$,
\item[Convolution-Multiplication] 
 \begin{equation}
 conv(\vec{x},\vec{y}) \Leftrightarrow \vec{X}*\vec{Y}
 \label{eq:conv-mul}
 \end{equation}
where $\vec{X}*\vec{Y}$ is the element-to-element vector multiplication
of two vectors $\vec{X}$ and $\vec{Y}$ , and
\item[Parseval's Relation] 
\begin{equation}
E(\vec{x}) = E(\vec{X}).
\label{eq:parseval}
\end{equation}
\end{description}
Using Parseval's relation, it is easy to show that the Euclidean
distance between two signals in the time domain is the same
as their distance in the frequency domain.
\begin{equation}
D(\vec{x},\vec{y}) = (E(\vec{x} - \vec{y}))^{1/2} = 
(E(\vec{X} - \vec{Y}))^{1/2} = D(\vec{X},\vec{Y})
\label{eq:pars-eucl}
\end{equation}

A nice property of the DFT is that for a large family of sequences it
concentrates the energy in the first few coefficients. Thus using the first
few coefficients for indexing introduces few false hits, and no false
dismissals. 

\subsection{Related Work}
There has been some work
on access methods for similarity queries. For example, Agrawal
et al. \cite{AgFaSw93} propose an efficient index structure to
retrieve time sequences similar to a given one. They map time sequences
into the frequency domain using the Fourier transform and keep the first few
coefficients in the index.
Two sequences are considered similar if their Euclidean distance is
less than a user-defined threshold. One difficulty with this approach
is that the user has no control over the meaning of similarity.

Jagadish et al.\cite{JMM95} develop a domain-independent
framework to pose similarity queries on a database.
The framework has three components: a {\it pattern
language} {\em P}, a {\it transformation rule language} {\em T}, and
a {\it query language} {\em L}. An expression in {\em P} specifies a
set of data objects.
An object {\it A} is considered similar to an object {\it B}, if {\it B}
can be reduced to it by a sequence of transformations defined in {\em T}.
The query language proposed in the paper is an extension of relational 
calculus with
predicates that test whether an object $A$ can be transformed into
a member of the set of objects described by expression $e$ using the
transformation $t$, at a cost bounded by $c$.
A specialization of this work to real-valued sequences where
the search is performed over sequence signatures 
instead of the original sequences is described in \cite{FJMM95}.

In this paper, we describe an efficient implementation of
a special case of
\cite{JMM95}
for time-series data.
We only study the trivial pattern language where a pattern expression
specifies either a given constant data object, or every object in the
database. 
Given an object $o$, a pattern expression $e$ that denotes a set of objects,
and a transformation $t$, the expression
$t(e)$\footnote{This is called $e \approx t$ in \cite{JMM95}.}
denotes the set
of all objects that can be obtained by applying $t$ to every member
of the set defined by $e$. We consider three kinds of queries:
range queries, all-pair queries, and nearest neighbor queries, and
we allow our transformations to be used in those queries.

We show how to use the indexing method in
\cite{AgFaSw93} to test for
similarity under a general class of transformations.

Our work generalizes 
Goldin et al. \cite{GoKa95} where transformations are limited
to {\it shifts} and positive {\it scales}. Our extension allows
shifts and scales in every dimension of a multidimensional
feature space, as well as more complex transformations such as moving average.
In addition, we drop the restriction to positive scales.
Some advantages of these extensions are shown within the next section.

There are other related works on time series data.  
Agrawal et al. \cite{shape95} describe a pattern language called SDL 
to encode queries about ``shapes'' found in time series. 
The language allows a kind of blurry matching where the user
specifies the overall shape instead of the specific details, but
it does not support any operations or transformations on the time series.
A query language for time series data in the stock market domain is 
described in \cite{mimsy}. The language is built on top of 
CORAL \cite{RSS92}, and every query is translated into a sequence of
CORAL rules. 

\section{Examples from Stock Data Analysis}
\label{sec:examples}
In this section we demonstrate how our transformations
can be used to eliminate noise or short-term fluctuations and
shift or scale the data before computing Euclidean distances.
We use
three examples from real stock data. The data was obtained from
the FTP site ``ftp.ai.mit.edu/pub/stocks/results/''.
 
\begin{figure*}[htbp]
\centering
\centerline{\psfig{figure=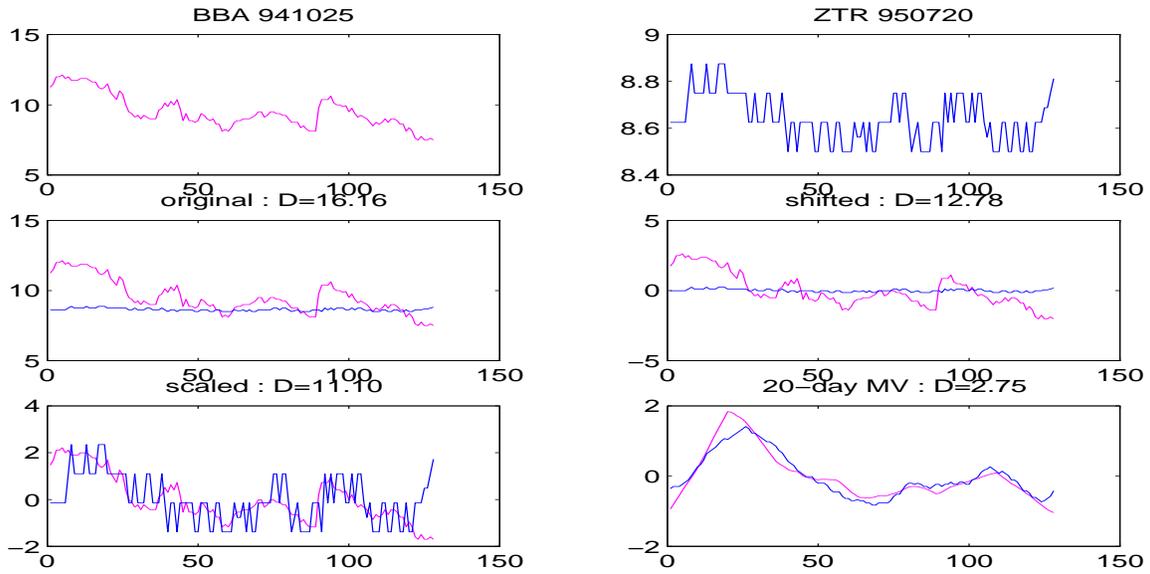,height=3in,width=6in}}
\caption{From left to right, top to bottom: the daily closing price
for {\it The Bombay Co. (BBA)} starting from ``94/10/25'' for 128 days,
the daily closing price for {\it Zweig Total Return Fund Inc. (ZTR)} starting
from ``95/07/20'' for 128 days, the two stocks put together, both shifted,
both scaled, and the 20-day moving average (D denotes the Euclidean distance)}
\label{fig:ex-bba-ztr}
\end{figure*}

\begin{example}{\rm
\label{example:21}
Figure~\ref{fig:ex-bba-ztr} shows the
daily closing price for {\it The Bombay Co.} (BBA) starting from 
October 25th, 1994 for
128 days, and that for {\it Zweig Total Return Fund Inc.} (ZTR) 
starting from July 20th, 1995 for 128 days.
The Euclidean distance between two series is 16.16. The mean for BBA
is 9.51, and the mean for ZTR is 8.64. If we shift the mean of both 
series to zero, i.e, subtract the mean of each series from everyday 
closing price, the Euclidean distance reduces to 12.78. The closing price
of ZTR fluctuates in a smaller range than that of BBA; the standard
deviation for ZTR is 0.10 while the standard deviation for BBA
is 1.18. We scale both series by the inverse of their standard deviation.
The resulting series in \cite{GoKa95} are called {\it normal forms} of 
the original series. Thus given any sequence $\vec{s}$, 
sequence $\vec{s^\prime}$ is the normal form of $\vec{s}$ if
\begin{equation}
s^\prime_i = \frac{s_i - mean(\vec{s})}{std(\vec{s})} \ \ \
for\ i=1,\cdots,length(\vec{s})
\end{equation}
Figure~\ref{fig:ex-bba-ztr} shows that the Euclidean distance between
the normal forms of two series is still 11.10; time series ZTR is
more volatile than BBA. To smooth out short term fluctuations, we take the
20-day moving average of the two series. The Euclidean distance drops to
2.75.
}\end{example}

\begin{figure*}[htbp]
\centering
\centerline{\psfig{figure=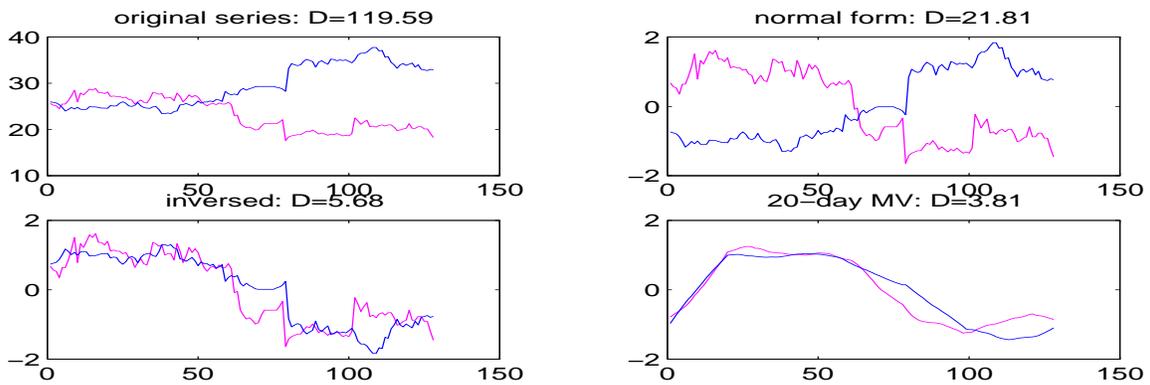,height=2in,width=6in}}
\caption{From left to right, top to bottom: the daily closing prices
for {\it Circuit City Stores Inc. (CC)} (marked by dotted lines) and 
{\it Varian Associates Inc. (VAR)} (marked by solid lines)
both starting from ``93/08/30'' for 128 days,
their normal forms, VAR reversed, and the 20-day moving averages of both
series (D denotes the Euclidean distance)}
\label{fig:ex-cc-var}
\end{figure*}

\begin{example}{\rm
\label{example:22}
This example shows how we can identify series that have opposite
price movements. Figure~\ref{fig:ex-cc-var} shows the daily closing 
price for {\it Circuit City Stores Inc.} (CC) (marked by dotted lines)
and the daily closing price for {\it Varian Associates Inc.} (VAR) (marked
by solid lines) both starting from August 30, 1993 for 128 days.
As Figure~\ref{fig:ex-cc-var} shows, the two series have a reverse movement;
when the price for CC goes up, the price for VAR goes down and vice versa.
The Euclidean distance between two series is 119.59. We transform
both series to their normal form, and the Euclidean distance becomes 21.81.
If we reverse the time series of VAR, i.e., multiply everyday closing
price by -1, and then take the 20-day moving average of both series,
the Euclidean distance will be 3.81. 
}\end{example}

One might think that applying these transformations, any two series
can be made similar.
The next example shows this is not the case.
Given three operations: {\it shift}, {\it scale},
and {\it 20-day moving average}, we can use {\it shift} and 
{\it scale} to transform two series to their normal forms. 
We can smooth the normal form series using {\it 20-day moving average}. 
Each one of these operations may reduce the distance between two series, 
but two series that have dissimilar trends still look different.
It is obvious that if we keep taking the moving average, two
series eventually will be the same, i.e., two flat straight lines.
However, we assume,
following \cite{JMM95}, that
each operation has a cost, and we are limited
by an upper bound on the total cost. This upper bound, for example,
could be proportional to the Euclidean distance between the two original
series.

\begin{figure*}[htbp]
\centering
\centerline{\psfig{figure=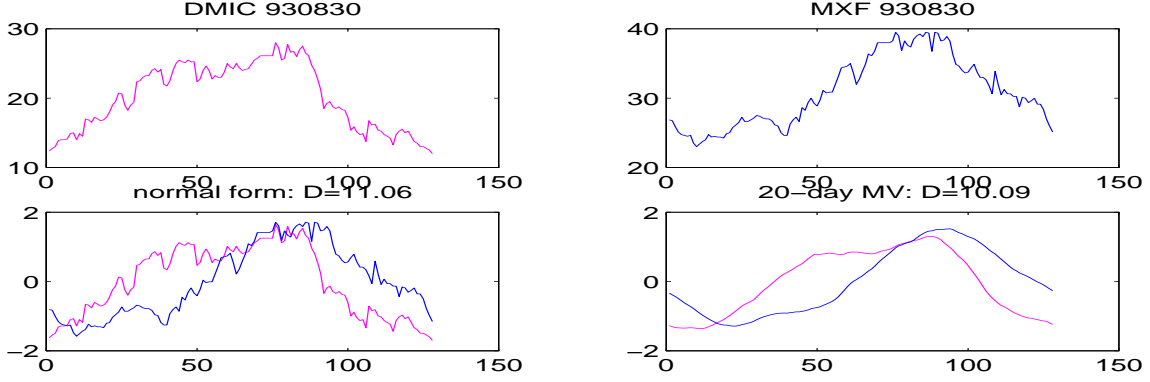,height=2in,width=6in}}
\caption{From left to right, top to bottom: the daily closing price
of {\it Digital Microwave Corp. (DMIC)} starting from ``93/08/30''
for 128 days, the daily closing price of {\it The Mexico Fund Inc. (MXF)}
for the same period, their normal forms, and the 20-day moving averages
of both series (D denotes the Euclidean distance)}
\label{fig:dmic-mxf}
\end{figure*}

\begin{example}{\rm
\label{example:23}
Figure~\ref{fig:dmic-mxf} shows the daily closing prices
of {\it Digital Microwave Corp. (DMIC)} and 
{\it The Mexico Fund Inc. (MXF)} both starting from August 30,1993
for 128 days. The Euclidean distance between the normal forms of
two series is 11.06. The Euclidean distance after taking the
20-day moving average becomes 10.09. The Euclidean distances after
taking the second and the third 20-day moving average are respectively
9.63 and 9.22. The Euclidean distance even after taking the 10th moving 
average is still 6.57.
}\end{example}

In the next section we encode the transformations discussed here
in a query language.

\section{Similarity Queries}
\label{sec:simq}

We consider an object to be a point in a multidimensional space (md-space). 
For non-point objects, we assume there is a mapping function that 
maps every object to a point in the md-space. 
Such a function is developed in many domains where multidimensional indexing
has been used. For example,
Fourier transform for time-series \cite{AgFaSw93}, and minimum 
bounding rectangle for shapes \cite{Jag91} are some instances of the mapping 
function.

\begin{figure}[htbp]
\psfig{figure=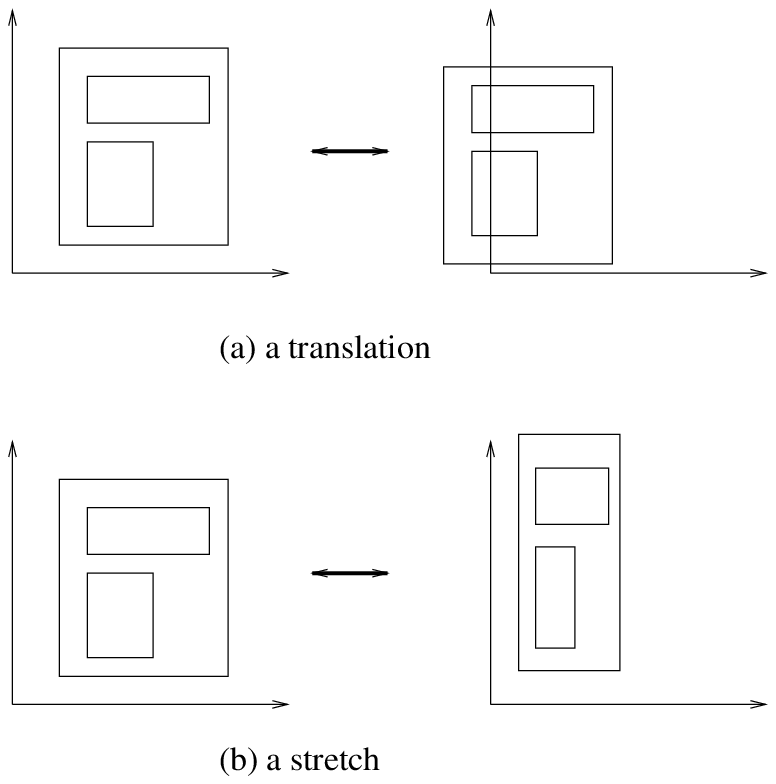,height=3in,width=3in}
\caption{}
\label{fig:trans}
\end{figure}

A transformation in an n-dimensional space, denoted by 
$(\vec{a},\vec{b})$, is a pair of $n \times 1$ vectors where
$\vec{a}$ specifies a stretch and $\vec{b}$ represents a translation
(Figure~\ref{fig:trans}). The transformation
$(\vec{a},\vec{b})$ applied to a point $\vec{x}$ in some space, maps $\vec{x}$
to $\vec{a}*\vec{x} + \vec{b}$ which is a point in the same space.
Transformations may be associated with costs. Given a set of transformations 
$t$, and the cost of applying each transformation, a measure of distance
(dissimilarity) between two objects can be defined as follows:
\begin{equation}
D(\vec{x},\vec{y}) = min \left\{
                     \begin{array}{l}
                     D_0(\vec{x},\vec{y})\\
                     min_{T\in t}(cost(T) + D(T(\vec{x}),\vec{y}))\\
                     min_{T\in t}(cost(T) + D(\vec{x},T(\vec{y})))\\
                     min_{T_1,T_2\in t}(cost(T_1) + cost(T_2)\\ 
                              \hspace*{2em} + D(T_1(\vec{x}),T_2(\vec{y})))\\
                     \end{array}
                     \right.
\end{equation}
where $D_0(\vec{x},\vec{y})$ is the Euclidean distance between $\vec{x}$
and $\vec{y}$.
We use $T(\vec{x})$ to denote ``transformation $T$ applied
to a point $\vec{x}$'' and 
$ T(r)$ to denote 
``transformation $T$ applied to a relation $r$''. The former
returns a point while the latter returns a relation.
We assume relations are unary, that is, they are simply sets of
sequences; in practice of course they may have other attributes,
such as source of the data, time period covered, etc.

In the domain of time series data, both objects and transformations
can be vectors of complex numbers, so we need to extend 
transformations to complex numbers. On the other hand, we want
to make sure that this extension still keeps the main properties 
we are interested in. The following definition describes these properties.

\begin{definition}
A transformation $T$ in a multidimensional space $S$ is {\em safe}
if $T$ maps every rectangle $R$ in $S$ to a
rectangle $R^\prime$ in $S$, every point inside $R$ to a point inside
$R^\prime$, and every point outside $R$ to a point outside $R^\prime$.
\end{definition}

\begin{theorem}
\label{theorem:trans}
Transformation $T=(\vec{a},\vec{b})$ is safe
if $\vec{a}$ and $\vec{b}$ are
chosen to be vectors of real numbers.
\end{theorem}

{\it Proof (sketch):} Transformation $T$ here is the composition of a
stretch and a translation in every dimension. Thus $T$ is safe.
\qed

In the next section, we study the safety condition for complex numbers
in more detail.

\subsection{Transformations on Time Series}
\label{sec:complexN}
We consider a time series to be a point in a multidimensional feature
space. We have chosen the first $k$ Fourier coefficient of a time series as
our features. 
The reason for choosing DFT is mainly because: (a) DFT concentrates 
the energy in the first few coefficients, so those coefficients can make 
a key for indexing purposes;
(b) as we remarked in Section~\ref{sec:DFT}, it is known that the 
Euclidean distance is unchanged under the DFT.
Since a Fourier coefficient, in general, is a complex number,
we need a representation for complex numbers in our feature space.
One possibility is to decompose a complex number into its
real and imaginary components, and map each component to a dimension.
For a given set of $k$ features, we represent it with a point in a
$2k$-dimensional space. We denote the space built using this
representation by $S_{rect}$.
An alternative representation is to decompose a complex number into its
components in the polar coordinate system.
A complex number in the polar coordinate system
is represented by a magnitude
and a phase angle. We denote the space built using this representation
by $S_{pol}$.
We use $Re(x)$, $Im(x)$, $Abs(x)$, and $Angle(x)$ to denote
respectively the real, the imaginary, the magnitude,
and the phase angle of a complex number $x$.
Now we need to show that the transformations described for time series
data in previous sections are safe.

\begin{theorem}
\label{theorem:rect}
Let $\vec{a}$ be a vector of real numbers, and $\vec{b}$ be a vector
of complex numbers;
the transformation $T=(\vec{a},\vec{b})$ is safe with respect to
$S_{rect}$.
\end{theorem}

{\it Proof:} We need to prove that $T$ maps every rectangle $R$ in
the space to a rectangle $R^\prime$, all interior points of $R$ to interior
points of $R^\prime$, and all exterior points of $R$ to exterior points of
$R^\prime$.
Without loss of generality, we assume dimensions $2i-1$ and
$2i$ (for $i=1,\cdots,k$) are respectively used for real and imaginary
components of feature $i$.
Suppose $\vec{xc}$ is a $k$-dimensional vector of complex numbers
and $\vec{x}$,
a $2k$-dimensional vector,
is its representation in $S_{rect}$.
We have $xc_i = x_{2i-1} + x_{2i}j$ for $i=1,\cdots,k$.
If we apply transformation $T$ to $\vec{xc}$, we get
$\vec{xc^\prime} = T(\vec{xc}) = \vec{a} * \vec{xc} + \vec{b}$.
We can rewrite this as follows:
\begin{eqnarray*}
xc_i^\prime & = & a_i * (x_{2i-1} + x_{2i}j) + (Re(b_i) + Im(b_i)j) \\
            & = & (a_i * x_{2i-1} + Re(b_i)) + (a_i * x_{2i} + Im(b_i))j
\end{eqnarray*}
for $i=1,\cdots,k$.
If we map the resulting vector to a point $x^\prime$ in $S_{rect}$, we get
$x_{2i-1}^\prime = a_i * x_{2i-1} + Re(b_i)$ and
$x_{2i}^\prime = a_i * x_{2i} + Im(b_i)$ for $i=1,\cdots,k$.
This transformation can be rewritten as $T^\prime = (\vec{c},\vec{d})$
where $c_{2i-1} = c_{2i} = a_i$, $d_{2i-1} = Re(b_i)$, 
and $d_{2i} = Im(b_i)$ for $i=1,\cdots,k$. Since $\vec{c}$ and $\vec{d}$
are vectors of real numbers, 
the rest follows from Theorem~\ref{theorem:trans}. \qed

On the other hand, Theorem~\ref{theorem:rect} does not hold if $\vec{a}$
is chosen to be a vector of complex numbers.
For example consider a two dimensional rectangle with point $p=-5 -5j$
as its lower left corner and point $q= 5 + 5j$ as its upper right corner,
and $r = -2 + 2j$ as a point inside the rectangle.
If we multiply the complex numbers representing the three points
by $s = 2 - 3j$, the transformed
rectangle built on points $p*s = -25 + 5j$ and $q*s = 25 - 5j$
does not have point $r*s = 2 + 10j$ inside!

\begin{theorem}
\label{theorem:polar}
Let $\vec{a}$ be a vector of complex numbers, and $\vec{b}$ be a zero
vector ($\vec{b} = \vec{0}$);
the transformation $T=(\vec{a},\vec{b})$ is safe with respect to
$S_{pol}$.
\end{theorem}

{\it Proof:}
Without loss of generality, we assume dimensions $2i-1$ and
$2i$ (for $i=1,\cdots,k$) are respectively used for
magnitude and phase angle of feature $i$.
Suppose $\vec{xc}$ is a $k$-dimensional vector of complex numbers
and $\vec{x}$ is its coordinate in $S_{pol}$.
We have $xc_i = x_{2i-1} e^{x_{2i}j}$ for $i=1,\cdots,k$.
If we apply transformation $T$ to $\vec{xc}$, we get
$\vec{xc^\prime} = T(\vec{xc}) = \vec{a} * \vec{xc} + \vec{b}$.
We can rewrite this as follows:
\begin{eqnarray*}
xc_i^\prime & = & Abs(a_i) e^{Angle(a_i)j} * x_{2i-1} e^{x_{2i}j} + 0 \\
            & = & (Abs(a_i)*x_{2i-1}) e^{(x_{2i}+Angle(a_i))j}
\end{eqnarray*}
for $i=1,\cdots,k$.
If we map the resulting vector to a point $x^\prime$ in $S_{pol}$, we get
$x_{2i-1}^\prime = Abs(a_i)*x_{2i-1}$ and
$x_{2i}^\prime = x_{2i}+Angle(a_i)$ for $i=1,\cdots,k$.
This transformation can be rewritten as $T^\prime = (\vec{c},\vec{d})$
where $c_{2i-1} = Abs(a_i)$, $d_{2i-1} = 0$, $c_{2i} = 1$, and
$d_{2i} = Angle(a_i)$for $i=1,\cdots,k$. Since $\vec{c}$ and $\vec{d}$
are vectors of real numbers, 
the rest follows from Theorem~\ref{theorem:trans}. \qed

\begin{figure}[htbp]
\psfig{figure=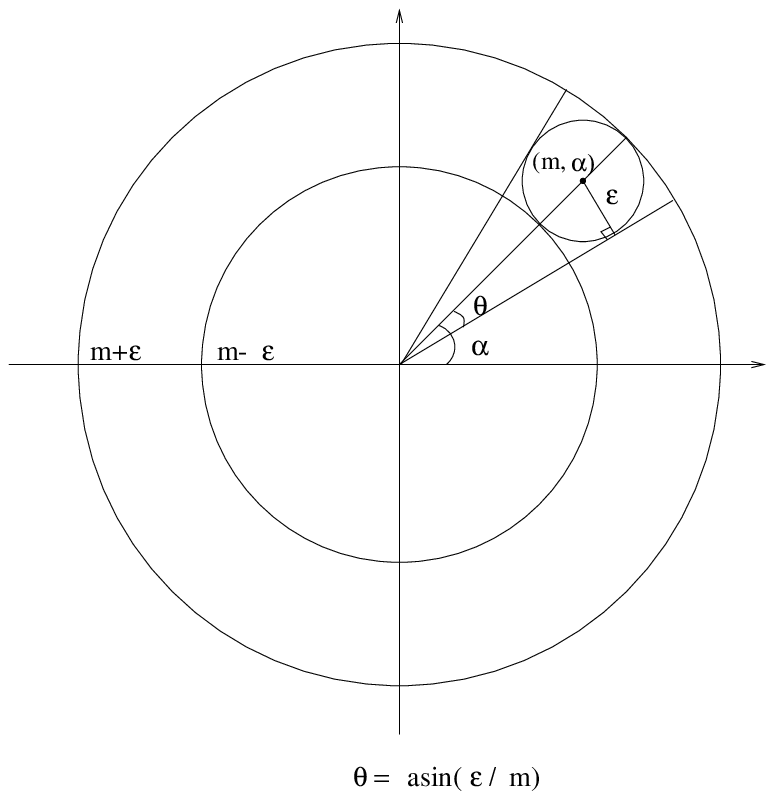,height=2in,width=2in}
\caption{Minimum bounding rectangle in the polar coordinate system}
\label{fig:threspolar}
\end{figure}

Given a query point $\vec{q}$ in a $2k$ dimensional space and a
threshold $\epsilon$, we need to build a search rectangle. 
A search rectangle is the minimum bounding rectangle that contains 
all points within the Euclidean distance $\epsilon$ from $\vec{q}$.
It is straightforward in the rectangular coordinate system; the minimum 
bounding rectangle is $(q_i - \epsilon, q_i + \epsilon)$ for $i=1,\cdots,2k$.
The minimum bounding rectangle for a complex number $m e^{\alpha j}$
in the polar coordinate system is demonstrated in Figure~\ref{fig:threspolar}.
The magnitude is in the range from $m-\epsilon$ to $m+\epsilon$, and
the angle is in the range from $\alpha - asin(\frac{\epsilon}{m})$ to
$\alpha + asin(\frac{\epsilon}{m})$ where $asin$ denotes the arc 
sinus of an angle.
If we again assume dimensions $2i-1$ and
$2i$ (for $i=1,\cdots,k$) are respectively used for
magnitude and phase angle of feature $i$,
then the minimum bounding rectangle for $\vec{q}$ in the polar coordinate
will be $(q_i - \epsilon, q_i + \epsilon)$ for $i=1,3,5, \cdots, 2k-1$ and
$(q_i - asin(\frac{\epsilon}{q_{i-1}}), q_i + asin(\frac{\epsilon}{q_{i-1}}))$
for $i=2,4,6, \cdots, 2k$.

\subsection{Using Transformations to Express Similarities}
To gain some insight into the transformations, we formalize
the notion of similarity expressed in Example~\ref{example:1}.
Time series $\vec{s_1}$ is considered similar to $\vec{s_2}$ because
their 3-day moving averages look the same, so we need to formulate 
the 3-day moving average in our transformation language.
For simplicity, in our examples we assign a cost of zero to all 
transformations.
Let us denote the Fourier transform of $\vec{s_1}$ by $\vec{S_1}$,
the Fourier transform of $\vec{s_2}$ by $\vec{S_2}$, and
the Fourier transform of $\vec{m_3} = (\frac{1}{3},\frac{1}{3},
\frac{1}{3},0,0,0,0,0,0,0,0,0,0,0,0)$ by $\vec{M_3}$.
Now consider the transformation $T_{mavg3} = (\vec{M_3},\vec{0})$ where
$\vec{0}$ is a zero vector of the same size as $\vec{M_3}$. 
If we apply the transformation $T_{mavg3}$ 
to $\vec{S_1}$, i.e., 
\[
T_{mavg3}(\vec{S_1}) = \vec{S_1} * \vec{M_3} + \vec{0} = \vec{S_1} * \vec{M_3}
\]
we get the 3-day moving average of $\vec{s_1}$ in the frequency domain.
If we transform the right hand side back to the time domain using the 
convolution-multiplication relation (Equation~\ref{eq:conv-mul}), we get 
$T_{mavg3}(\vec{s_1}) = conv(\vec{s_1},\vec{m3})$ which is the 3-day moving 
average of $\vec{s_1}$ in the time domain. The same transformation can
be applied to $\vec{s_2}$ to compute its 3-day moving average.

The notion of similarity can be expressed in a query by the proper choice of 
transformations. For example, the $m$-day moving average of a series
of length $n$ generally can be expressed by $T_{mavg}=(\vec{a},\vec{0})$ where 
\begin{equation}
\vec{a}=(\underbrace{\underbrace{w_1 = 1/m, w_2 = 1/m, \cdots, w_m = 1/m}_{m}, 
                                 0, 0, \cdots, 0}_{n})
\label{eq:mavg-A-m}
\end{equation}
and $\vec{0}$ is a zero vector of size $n$.
Transformation $T_{mavg}$ may be applied several times to get successive
moving averages. The weights $w_1, \cdots, w_m$ are not necessarily equal. 
For trend prediction purposes, for example, the weights at the end are 
usually chosen to be higher than those at the beginning. Whereas for normal 
smoothing purposes, weights are equal, or those at the center are larger than 
those at the endpoints.

To give another example of the transformations,
we formulate the transformation used to reverse a time series in
Example~\ref{example:22}. Let $T_{rev}=(\vec{a},\vec{0})$ where 
$a_i=-1$ for $i=1,\cdots,128$ and $\vec{0}$ is a zero vector of size 128. 
Now consider a time series $\vec{s}$ and its Fourier transform $\vec{S}$.
Transformation $T_{rev}$ applied to $\vec{S}$ gives
\[
T_{rev}(\vec{S}) = \vec{a} * \vec{S} + \vec{0} = - \vec{S}.
\]
If we transform both sides into the
time domain using Equation~\ref{eq:linear},
we get $T_{rev}(\vec{s}) = -\vec{s}$. That is, the daily closing price
is multiplied by $-1$. 

Transformation $T_{rev}$ can be used to obtain all the pairs of series that
move in opposite directions. This can be formulated in our query language
for a given relation $r$ as a
spatial
join between $r$ and 
$T_{rev}(r)$.

Transformations can also be defined to stretch the time dimension 
(Example~\ref{example:2}). Details of this transformation are given 
in Appendix~\ref{appendix-texp}. In the next section, we discuss an indexing
technique for similarity queries.

\section{Indexing of Similarity Queries}
\label{sec:index}
In this section we describe a fast query processing method for
similarity queries. 
We assume a multidimensional index is available, and we take
an advantage of that in our query processing.
Because of the dominant use of the R-tree family in multidimensional
indexing, we describe our approach for R-tree indexes.

Given an R-tree index $I$ in a multidimensional space $S$ over a data set $D$, and any safe transformation $T$ in $S$, we give an algorithm to construct an R-tree index $I^\prime$ for
$T(D)$.

\begin{algorithm}{\bf :}  {\rm
\label{RTconstruct}
For every node $n$\\
\[
n = ((MBR_1,\ pointer_1), \cdots, (MBR_m,\ pointer_m))
\]
in $I$, we construct a node $n^\prime$
\begin{eqnarray*}
n^\prime & = & T(n) \\
         & = & ((MBR_1^\prime,\ pointer_1^\prime), \cdots, 
                (MBR_m,\ pointer_m^\prime))
\end{eqnarray*}
in $I^\prime$ such that $MBR_i^\prime = T(MBR_i)$, and $pointer_i^\prime$
is a pointer to $T(n_i)$ where $n_i$ is the child node (or the data tuple
when $n$ is a leaf node) pointed by $pointer_i$ (for $i=1,\cdots,m$).
We assumed $T$ is a safe transformation, therefore $MBR_i^\prime$
is a bounding rectangle for all rectangles of the child node (or the
data tuple) $T(n_i)$.
The construction stops when every node in $I$ is mapped to a
node in $I^\prime$. 
}\end{algorithm}

There are many possible R-tree indexes on $T(D)$, each 
with a different performance. Our experiments show that the index
we build here has a similar performance to that of the original index.
The main observation here is that for a given index $I$ and transformation
$T$, index $I^\prime$ can be built on the fly without having much impact
on the performance of the search.
This allows us to use one index for many transformations.

An indexing method for time series data is described in \cite{AgFaSw93}.
This method requires a cut-off point
for the number of Fourier coefficients kept in the index.
We denote this cut-off point by $k$ and call the index built on
the first $k$ Fourier coefficients `{\it k}-index'.

To demonstrate the query processing algorithm, we use a more general form of
Example~\ref{example:1} throughout this section.
We have seen in Section~\ref{sec:simq} that the $m$-day moving average of
a series can be expressed by $T_{mavg} = (\vec{a},\vec{0})$ where $\vec{a}$
is a vector of complex numbers.
Due to Theorem~\ref{theorem:polar}, $T_{mavg}$ is safe if we represent
complex numbers in the polar coordinate.
\begin{quote}
{\bf Query:} Given a pattern expression $e$, a safe transformation $T$, an 
object $\vec{q}$, and a threshold $\epsilon$, find all objects 
$\vec{o} \in T(e)$ such that 
the Euclidean distance $D(\vec{o},\vec{q}) < \epsilon$.
\end{quote}

If the pattern expression $e$ denotes only one object $\vec{o_1}$, we simply 
apply $T$ to $\vec{o_1}$. Object $\vec{o_1}$ is in the answer set 
if $D(\vec{o_1},\vec{q}) < \epsilon$.
Now suppose the expression $e$ denotes all objects in the relation.
A naive evaluation requires reading the whole relation, applying $T$
to every object, and choosing every object $\vec{o}$ such that
$D(\vec{o},\vec{q}) < \epsilon$.
This is a costly process.
A better approach is to use Algorithm~\ref{RTconstruct} to construct
a new index for transformed objects, and this new
index can be built on the fly during the search operation.
The search algorithm for the given range query is as follows:

\begin{algorithm}{\bf : }{\rm
Given a {\it k}-index whose root is N, a transformation $T$, a threshold 
$\epsilon$,
and a search point $\vec{q}$, apply $T$ to all points in the index and
find those whose distance from $\vec{q}$ becomes less than $\epsilon$.

\begin{enumerate}
\item {\bf Preprocessing:}

\begin{enumerate}
\item Transform $T$ and $\vec{q}$ into the frequency domain if
they are in the time domain. Let us denote the first $k$ Fourier 
coefficients of $T$ and $\vec{q}$ by $T_k$ and $\vec{q}_k$ respectively.
\item build a search rectangle $q_{rect}$ for $\vec{q}_k$ as 
described in Section~\ref{sec:complexN}.

\end{enumerate}

\item {\bf Search:}

\begin{enumerate}
\item If $N$ is not a leaf, apply $T$ to every (rectangle) entry of $N$
and check if the resulting rectangle overlaps $q_{rect}$. For all overlapping
entries, call {\bf Search} on the index whose root node is pointed to by the
overlapping entry.
\item If $N$ is a leaf, apply $T$ to every (point) entry of $N$
and check if the resulting point overlaps $q_{rect}$. If so, the entry
is a candidate. 
\end{enumerate}

\item {\bf Postprocessing:}

\begin{enumerate}
\item For every candidate point, check its full database record to determine
if its Euclidean distance is at most $\epsilon$ from $\vec{q}$. If so,
the entry is in  the answer set. 
\end{enumerate}
\end{enumerate}
}\end{algorithm}

Similarly all-pairs queries and nearest neighbor queries can be processed
efficiently using the index. For an all-pairs query, we do a spatial join
using the index. The only difference here is that we transform all
objects used in the join predicate before we compute the predicate. For 
example, the join predicate $a_i \cap b_j \neq \emptyset$ may
be changed to $T(a_i) \cap T(b_j) \neq \emptyset$ where $T$ is a
transformation and $a_i$ and $b_j$ are members of two spatial sets.
For a nearest neighbor query, the search starts from the root and 
proceeds down the tree. As we go down the tree, we apply $T$
to all entries of the node we visit. We can then use any kind
of metric (such as MINDIST or MINMAXDIST discussed in \cite{NNQ95}) 
for pruning the search.

The only thing left to show is that this search scheme 
used with a $k$-index misses no object from the answer set.

\begin{lemma} The $k$-index approach enhanced with 
transformations always returns a superset of the answer set.\\
\end{lemma}

{\it Proof:} Suppose we want to find all objects $\vec{x}$ in a relation
that are similar to a query object $\vec{q}$. Since transformations
are applied to series in the frequency domain, this can be written in
the frequency domain as follows:
\begin{equation}
D(T(\vec{X}),\vec{Q}) \leq \epsilon
\label{eq:lemma1:1}
\end{equation}
where $T=(\vec{A},\vec{B})$ is a transformation, $\epsilon$
is a user-defined threshold, and $\vec{X}$ and
$\vec{Q}$
are DFTs of
respectively $\vec{x}$ and
$\vec{q}$.
Applying $T$ to $\vec{X}$, we get 
\[
D(\vec{A} * \vec{X} + \vec{B},\vec{Q}) =
(\sum_{f=0}^{n-1} |A_f X_f + B_f - Q_f|^2 )^\frac{1}{2} \leq \epsilon
\]
If we keep only the first $k < n$ coefficients, we have
\begin{equation}
(\sum_{f=0}^{k-1} |A_f X_f + B_f - Q_f|^2 )^\frac{1}{2} \leq
(\sum_{f=0}^{n-1} |A_f X_f + B_f - Q_f|^2 )^\frac{1}{2}
\label{eq:lemma1:2}
\end{equation}
On the other hand, the equation
\begin{equation}
(\sum_{f=0}^{n-1} |A_f X_f + B_f - Q_f|^2 )^\frac{1}{2} \leq \epsilon
\label{eq:lemma1:3}
\end{equation}
holds for all objects in the answer set.
Equations (\ref{eq:lemma1:2},\ref{eq:lemma1:3}) imply
that 
\begin{equation}
(\sum_{f=0}^{k-1} |A_f X_f + B_f - Q_f|^2 )^\frac{1}{2} \leq \epsilon
\end{equation}
Therefore, keeping the first $k$ coefficients introduces no false
dismissals.
\qed

This is a generalization of the result of \cite{AgFaSw93} for $k$-index
enhanced with transformations.

\section{Experiments}
\label{sec:exper}
We implemented our method on top of Norbert Beckmann's Version 2 
implementation of the R*-tree \cite{BKSS90}. 
We ran experiments on both stock prices data obtained from the FTP site 
``ftp.ai.mit.edu/pub/stocks/results/'' and synthetic sequences. 
Each synthetic sequence $\vec{x} = [x_t]$ was a random sequence produced 
as follows:
\[
x_0 = y
\]
\[
x_1 = x_0 + z_1
\]
\[
\cdots
\]
\[
x_i = x_{i-1} + z_i
\]
where $y$ was a normally distributed random number in the range $[20, 99]$, and 
$z_t$ ($t=1, 2, \cdots $) was a random number in the range $[-4,4]$. 

We used the polar representation of complex numbers because
vector multiplication for time series data seemed to be more
important than vector addition, and due to Theorem~\ref{theorem:polar}
vector multiplication is safe with respect to $S_{pol}$. 
For every time series, we first transformed it to the normal form, and
then we found its Fourier coefficients. 
The reason for choosing the normal form was because both the average
and the standard deviation of a series could be stored in the index
as two separate dimensions, and despite using the polar representation,
we could still have simple shifts. 
Since the mean of a normal
form series is zero by definition, the first Fourier coefficient is
always zero, so we can throw it away. We mapped the mean and the standard
deviation of the original time series respectively to the first and the
second dimensions of the index. We also mapped the magnitude and
the phase angle of the second DFT term (computed for the normal form series)
respectively to the third and the fourth dimensions of the index, and   
the magnitude and the phase angle of the third DFT term
respectively to the fifth and the sixth dimensions.

\begin{figure}[htbp]
\psfig{figure=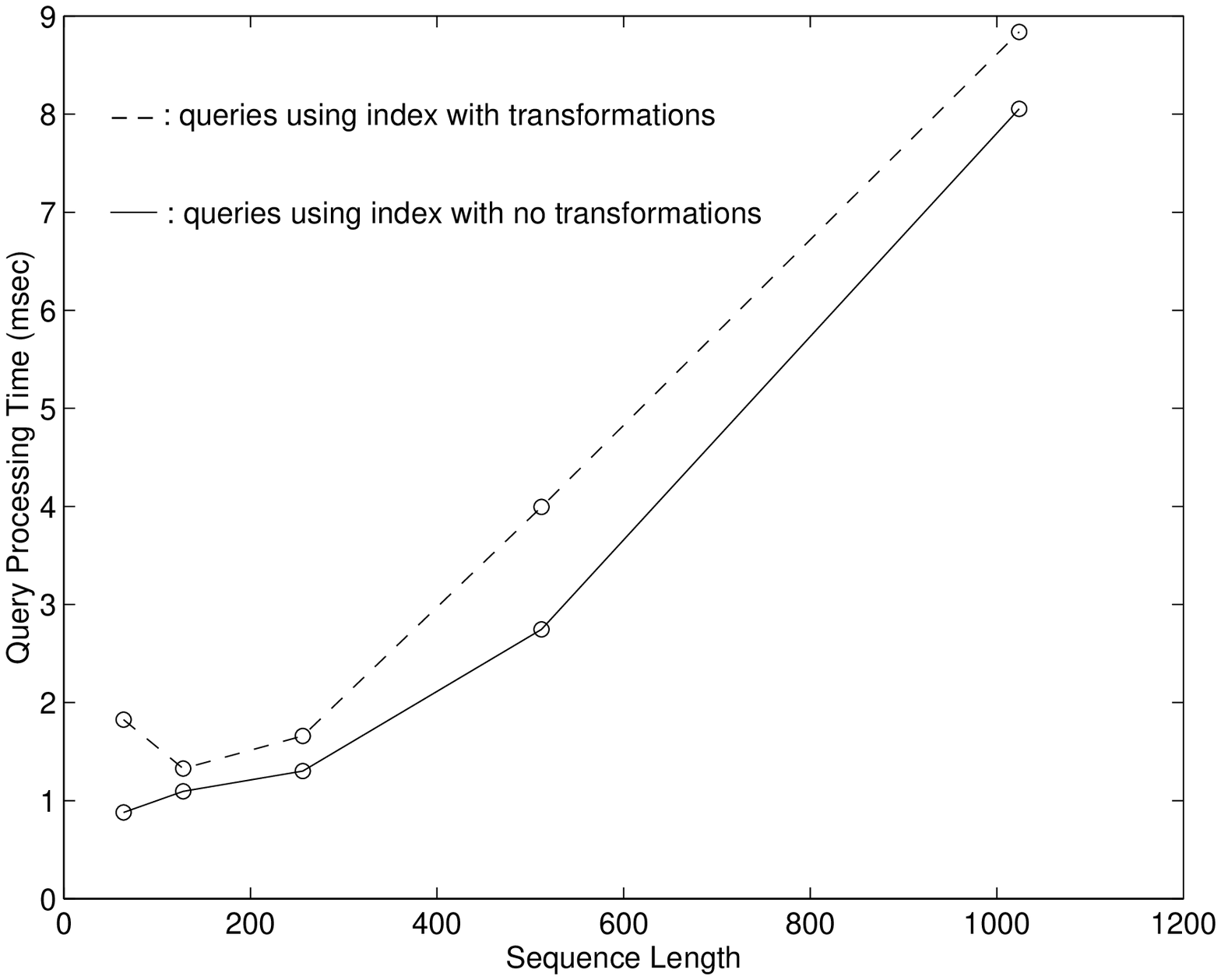,height=2in,width=3.05in}
\caption{time per query varying the sequence length}
\label{fig:graph1}
\end{figure}

Figure~\ref{fig:graph1} compares the execution time
for two kinds of queries: (a) a range query using transformations 
and (b) a range query that uses no transformations. We varied the length 
of the sequences from 64 to 1024 while we kept the number of sequences
fixed to 1,000. 
In order to have a precise comparison, the identity transformation 
$T_i = (\vec{I},\vec{0})$ was chosen such that $T_i(\vec{o})=\vec{o}$ for 
all objects $\vec{o}$ ($\vec{I}$ is a vector of 1's and $\vec{0}$ is a vector
of 0's). 
This made the two queries produce the same results. 
As Figure~\ref{fig:graph1} shows the difference between the two curves 
is only a constant. This constant is the 
CPU time spent for vector multiplication which is unavoidable. The number 
of disk accesses is the same in both cases.

\begin{figure}[htbp]
\psfig{figure=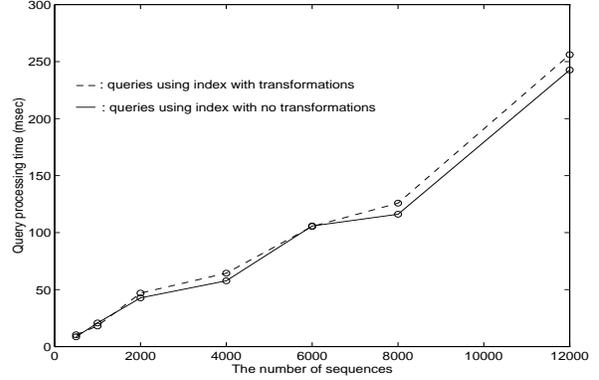,height=2in,width=3.05in}
\caption{time per query varying the number of sequences}
\label{fig:graph2-1}
\end{figure}

In the next experiment, we kept the sequence length fixed to 128 while
we varied the number of sequences from 500 to 12,000.
We used the identity transformation again for the same reason
described in the previous experiment.
As demonstrated in Figure~\ref{fig:graph2-1}, the result was the same.
Thus the index traversal for similarity queries does not deteriorate 
the performance of the index.

\begin{figure}[htbp]
\psfig{figure=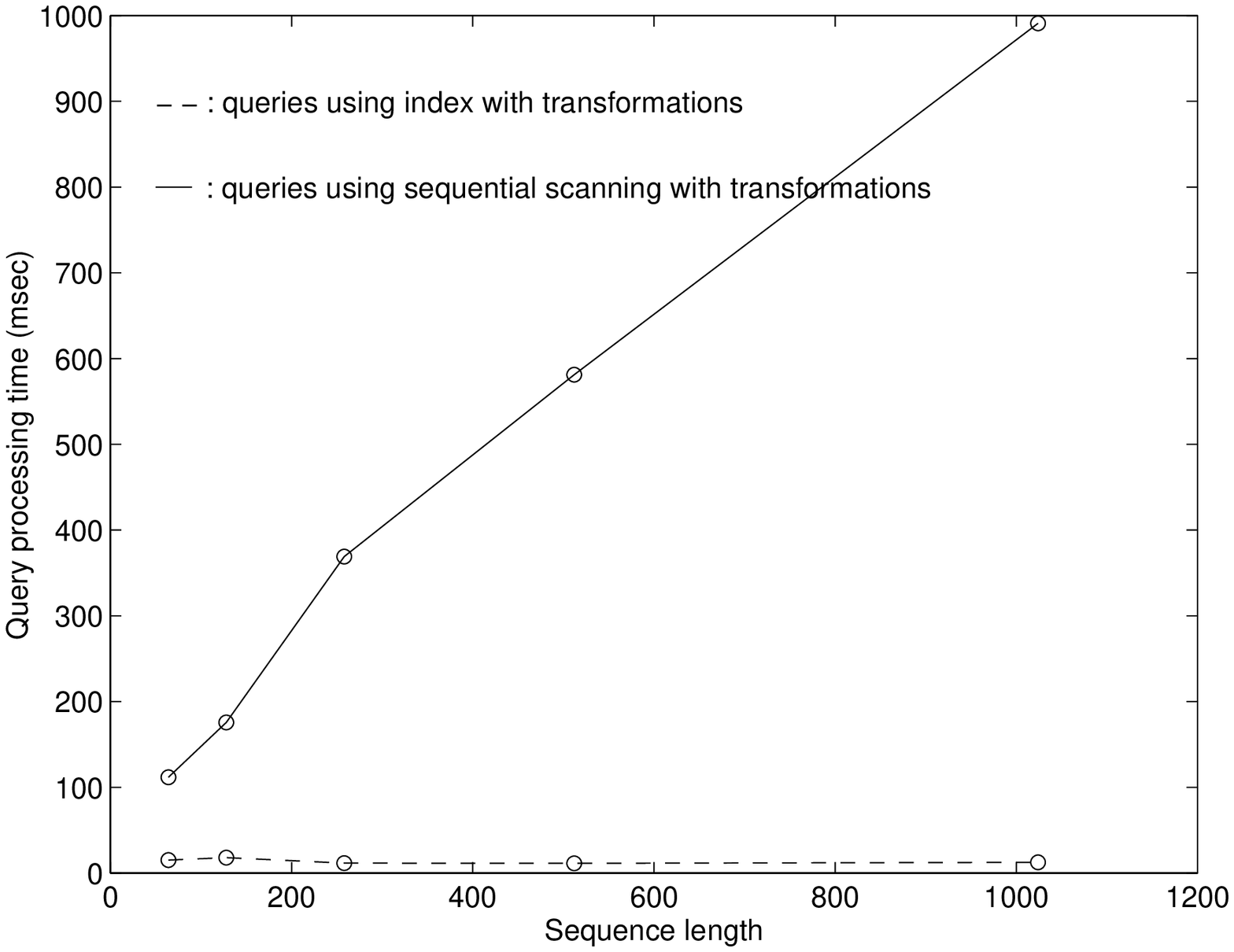,height=2in,width=3.05in}
\caption{time per query varying the sequence length}
\label{fig:graph1-2}
\end{figure}

\begin{figure}[htbp]
\psfig{figure=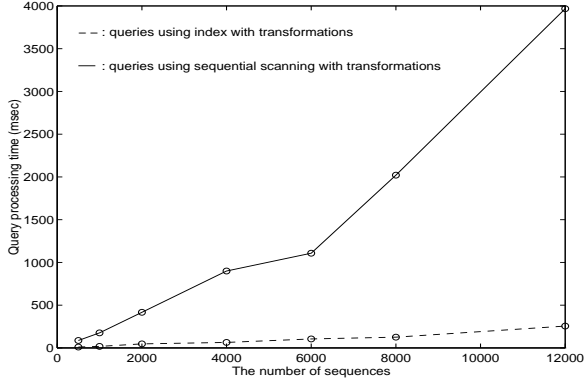,height=2in,width=3.05in}
\caption{time per query varying the number of sequences}
\label{fig:graph2-2}
\end{figure}

Figures~\ref{fig:graph1-2} and \ref{fig:graph2-2} compare
the execution time of our approach to sequential 
scanning. To have a good implementation for the sequential
scan, we stop the distance computation process as soon as
the distance exceeds $\epsilon$.
In addition, we do the sequential
scanning on the relation that stores the series in the frequency domain,
not the time domain.
Because each series in the frequency domain has its 
larger coefficients at the beginning, the distance computation
process can skip many sequences within the first few coefficients.
Both graphs show the superiority of our approach.

\begin{figure}[htbp]
\psfig{figure=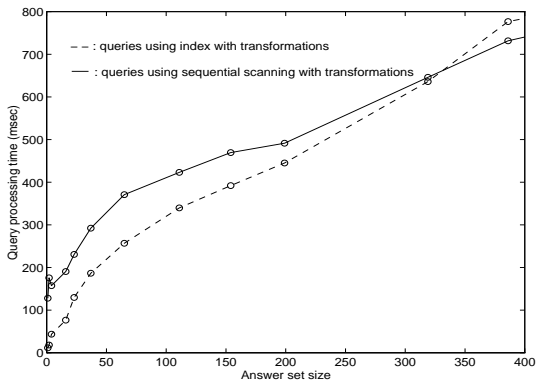,height=2in,width=3.05in}
\caption{time per query varying the size of the answer set}
\label{fig:graph3}
\end{figure}

In another experiment, we kept the number of sequences fixed to 1067, and
we also kept the sequence length fixed to 128. The experiment ran on the
real stock data. We varied the threshold 
so that the query gave us different numbers of time series in the
answer set. Figure~\ref{fig:graph3} shows that the index performs
better until the size of the answer set gets larger than 300 which
is almost one third of the size of the relation.  

Our last experiment was on a spatial self-join. 
In fact, all the time series used as examples in Section~\ref{sec:examples} 
were the results of this spatial join.
We did the test using the following methods: 
\begin{description}
\item[a] scan the relation of Fourier coefficients sequentially, and
compare every sequence $s$ to 
all the sequences that are after $s$ in the relation; 
the transformation $T_{mavg20}$ is 
applied to every sequence during the comparison; 
\item[b] do the sequential scanning as instructed in $a$, but 
stop the distance computation as soon as the distance exceeds $\epsilon$.
\item[c] scan the relation of Fourier coefficients sequentially, and
for every sequence build a search rectangle and pose it to the index
as a range query;
\item[d] do the spatial join as described in $c$, but in every index 
retrieval apply $T_{mavg20}$ to both the index and the search rectangles.
\end{description}
The experiment ran on a relation of stock prices data that
had 1067 time sequences, and  the length of each sequence was 128.
The result of the test is shown in table~\ref{table:jointest}.
\begin{table}[htbp]
\begin{tabular}{|l|l|l|} \hline
algorithm & time              & size of the \\ 
          & (min:sec.milisec) & answer set\\ \hline\hline
a & 20:36.323 & 12\\ \hline
b & 2:31.217 & 12\\ \hline
c & 0:10.139 & $3 \times 2 = 6$\\ \hline
d & 0:17.698 & $12 \times 2 = 24$\\ \hline
\end{tabular}
\caption{The result of the join}
\label{table:jointest}
\end{table}
Because of the implementation, $b$ is 10 times faster than $a$.
Methods $c$ and $d$ are 9 to 15 times faster 
than $b$ because of using the index, and $d$ is a bit 
slower than $c$ because of the use of a transformation and the
size of the answer set. 
The answer set of $d$ contains every pair twice, so it is 
twice the size
of $a$ and $b$. The size of the answer set for $c$ is smaller because it
does not use the transformation.

\section{Conclusions}
\label{sec:conclude}
We have proposed a class of transformations that can be used in a
query language to express similarity among objects. 
This class allows the expression of several practically important
notions of similarity, and queries using this class
can be
efficiently
implemented
on top of any R-tree index. 
One potential application which is emphasized in the paper
is stock data analysis, but we believe other domains can
also benefit. 
Our contributions can be summarized as follows:
\begin{itemize}
\item Formulation of moving average, time warping, and reversing
in our transformation language.
\item Implementation of similarity matching under these transformations
on top of an R-tree index. 
\end{itemize}

The experiments
show that execution time of our method
is almost the same as that of accessing the index with no
transformations; our method has much better performance than sequential
scanning, and the performance gets better by increasing 
both the number and the length of sequences.

We think the normal form of \cite{GoKa95}
is a useful representation for time series data, but
it allows only a small fraction of similarity queries.
Our similarity transformations allow more general queries, but for
simple shifting and scaling, the indexing method in \cite{GoKa95}
is faster because no transformation needs to be performed on the index.
Our indexing technique
can be easily built on top of \cite{GoKa95} as we did in our experiments,
allowing both simple shifts and scales and more general transformations
to be applied efficiently.

\section*{Acknowledgments}
We would like to thank Christos Faloutsos for his help in providing
us with stock data and his comments on a preliminary version
of this paper. This work was supported by the Natural Sciences and
Engineering Research Council of Canada and the Information Technology
Research Centre of Ontario.

\bibliography{ref}
\appendix
\section{Time Warping}
\label{appendix-texp}
Given the first $k \leq n$ Fourier coefficients of a time series $\vec{s}$ of
length $n$ and an integer $m \geq 1$, we can construct the first 
$k$ Fourier coefficients of time series $\vec{s^\prime}$ of length 
$m \times n$ using a transformation $T=(\vec{a},\vec{0})$ such that 
\begin{equation}
s^\prime_{mi} = s^\prime_{mi+1} = \cdots = s^\prime_{m(i+1) - 1} = s_i
\label{eq:app:te1}
\end{equation}
for $i=0,\cdots,n$.

Let $\vec{s} = (s_1, s_2, \cdots, s_n)$ be a time sequence, and 
$\vec{S} = (S_1, S_2, \cdots, S_n)$ be its DFT. Using Equation~\ref{eq:fft}, 
we can write
\begin{equation}
S_f = \frac{1}{\sqrt{n}}\sum_{t=0}^{n-1} s_t e^{\frac{-j2\pi tf}{n}} \ \ \
f=0,1, \cdots, n-1.
\label{eq:app:te2}
\end{equation}
We want to find a vector $\vec{a}$ such that
\begin{equation}
a_f * S_f = S^\prime_f \ \ \ for\ f=0,\cdots,k-1.
\label{eq:app:te3}
\end{equation}
where $S^\prime_f$ is the {\it f}th Fourier coefficient of $\vec{S^\prime}$.
Using Equation~\ref{eq:fft}, we can write $S^\prime_f$ as follows:
\[
S^\prime_f = \frac{1}{\sqrt{n}}\sum_{t=0}^{mn-1} 
             s^\prime_t e^{\frac{-j2\pi tf}{mn}}.
\]
This can be rewritten as
\[
S^\prime_f = \frac{1}{\sqrt{n}} (\sum_{t=0}^{m-1} 
             s^\prime_t e^{\frac{-j2\pi tf}{mn}} + 
             \sum_{t=m}^{2m-1} s^\prime_t e^{\frac{-j2\pi tf}{mn}} + \cdots
\]
\[
             + \sum_{t=(n-1)m}^{nm-1} s^\prime_t e^{\frac{-j2\pi tf}{mn}} ).
\]
If we rewrite all summations as $\sum_{t=0}^{m-1}$ and also use 
Equation~\ref{eq:app:te1}, we get
\[
S^\prime_f = \frac{1}{\sqrt{n}} (\sum_{t=0}^{m-1} 
             s_0 e^{\frac{-j2\pi tf}{mn}} +
             \sum_{t=0}^{m-1} s_1 e^{\frac{-j2\pi (t+m)f}{mn}} + \cdots
\]
\[
             + \sum_{t=0}^{m-1} s_{n-1} e^{\frac{-j2\pi (t+(n-1)m)f}{mn}} ).
\]
We can take $\sum_{t=0}^{m-1} e^{\frac{-j2\pi tf}{mn}}$ out of the
parenthesis; that gives
\[
S^\prime_f = \frac{1}{\sqrt{n}} \sum_{t=0}^{m-1} e^{\frac{-j2\pi tf}{mn}} (
             s_0 + s_1 e^{\frac{-j2\pi f}{n}} + \cdots +
             e^{\frac{-j2\pi (n-1)f}{n}}).
\]
Using Equation~\ref{eq:app:te2}, we can rewrite this equation as follows:
\[
S^\prime_f = \sum_{t=0}^{m-1} e^{\frac{-j2\pi tf}{mn}} S_f.
\]
Therefore, if we choose vector $\vec{a}$ as follows:
\begin{equation}
a_f = \sum_{t=0}^{m-1} e^{\frac{-j2\pi tf}{mn}} \ \ \ for\ f=0,\cdots,k-1,
\end{equation}
then the Equation~\ref{eq:app:te3} holds.

\end{document}